\documentclass[useAMS, usenatbib]{mn2e}
\usepackage[dvips]{graphicx}
\usepackage{amssymb}
\usepackage{multirow}
\input{epsf}

\newcommand{\ug}{$u$--$g$}
\newcommand{\gr}{$g$--$r$}
\newcommand{\ri}{$r$--$i$}
\newcommand{\iz}{$i$--$z$}
\newcommand{\Zz}{$Z$--$z$}
\newcommand{\ZY}{$Z$--$Y$}
\newcommand{\YJ}{$Y$--$J$}
\newcommand{\JH}{$J$--$H$}
\newcommand{\HK}{$H$--$K$}
\newcommand{\JJ}{$J$--$J$2}
\newcommand{\HH}{$H$--$H$2}
\newcommand{\KK}{$K$--$K$2}
\newcommand{\tf}{$T_{\mathrm {eff}}$}

\title[The UKIDSS Photometric System] {The UKIRT Infrared
Deep Sky Survey $Z$$Y$$J$$H$$K$ Photometric System: Passbands and
Synthetic Colours}

\author[Hewett et al.]
{P. C. Hewett,$^{1}$ S. J. Warren,$^{2}$ S. K. Leggett,$^{3}$
S. T. Hodgkin$^{1}$ \\
$^{1}$Institute of Astronomy, Madingley Road, Cambridge CB3 0HA\\
$^{2}$Blackett Laboratory, Imperial College of Science Technology
and Medicine, Prince Consort Rd, London SW7 2AZ \\
$^{3}$United Kingdom Infrared Telescope, Joint Astronomy Centre,
660 North A`ohoku Place, Hilo, HI 96720\\}

\date{Accepted
        Received
        in original form}

\begin{document}
\maketitle

\begin{abstract}

The UKIRT Infrared Deep Sky Survey is a set of five surveys of
complementary combinations of area, depth, and Galactic latitude,
which began in 2005 May.  The surveys use the UKIRT Wide Field Camera
(WFCAM), which has a solid angle of 0.21deg$^2$.  Here we introduce
and characterise the $Z$$Y$$J$$H$$K$ photometric system of the camera,
which covers the wavelength range $0.83-2.37\,\mu$m.  We synthesise
response functions for the five passbands, and compute colours in the
WFCAM, SDSS and 2MASS bands, for brown dwarfs, stars, galaxies and
quasars of different types.  We provide a recipe for others to compute
colours from their own spectra.  Calculations are presented in the
Vega system, and the computed offsets to the AB system are
provided, as well as colour equations between WFCAM filters and the
SDSS and 2MASS passbands.  We highlight the opportunities presented by
the new $Y$ filter at $0.97-1.07\,\mu$m for surveys for hypothetical Y
dwarfs (brown dwarfs cooler than T), and for quasars of very--high
redshift, $z>6.4$.

\end{abstract}


\section{Introduction}

The UKIRT Infrared Deep Sky Survey (UKIDSS, Lawrence et al.  2006, in
preparation) commenced on 2005 May 13, and is a set of five surveys of
complementary combinations of depthand area, employing the wavelength range
$0.83-2.37\,\mu$m in up to five filters $ZYJHK$, and extending over both high
and low Galactic latitude regions of the sky.  The new $Z$ band,
$0.84-0.93\,\mu$m, and the $Y$ band, $0.97-1.07\,\mu$m, introduced by Warren \&
Hewett (2002), are both characterised in this paper.  It is anticipated that the
surveys will take seven years to complete.  This paper characterises the
photometric system of the survey, and presents synthetic colours of a wide
variety of sources.  The paper is one of a set of five which provide the
reference technical documentation for UKIDSS.  The other four papers, described
below, are Casali et al.  (2006, in preparation), Lawrence et al.  (2006), Irwin
et al.  (2006, in preparation) and Hambly et al.  (2006, in preparation).

In this paper all quoted magnitudes are on the Vega system.  We use the
nomenclature $Z$, $Y$, $J$, $H$, $K$ for magnitudes in the five bands used by
UKIDSS; $J2$, $H2$, $K2$ for the bands of the 2Micron All Sky Survey (2MASS,
Cutri et al.  2003), and $u$, $g$, $r$, $i$, $z$ for the native bands of the
Sloan Digital Sky Survey (SDSS, York et al.  2000) 2.5m telescope.  Vega--based
magnitudes in the SDSS system have been calculated by assuming that Vega has
zero magnitude in the $u$, $g$, $r$, $i$ and $z$ passbands.  Appropriate
zero--point offsets to allow conversion to the {\em AB} system are given in
Section 4.6. Quoted survey depths correspond to $5\sigma$ significance for a
point source.

The survey instrument is the Wide Field Camera (WFCAM) on the United
Kingdom Infrared Telescope (UKIRT).  A detailed description of the
instrument is provided by Casali et al.  (2006).  The camera has four
Rockwell Hawaii-II $2048^2$ PACE arrays, with pixel scale
$0.4\arcsec$, giving a solid angle of 0.21$\,$deg$^2$ per exposure.
At the time of commissioning, 2004 November, the instrument {\em
\'{e}tendue}\footnote{product of telescope collecting area, and solid
angle of instrument field of view, sometimes called {\em grasp}} of
$2.38\,$m$^2\,$deg$^2$ was the largest of any near-infrared imager in
the world.  The Canada France Hawaii Telescope WIRCam instrument
(Puget et al.  2004) covers a solid angle of 0.11$\,$deg$^2$ per
exposure giving an {\em \'{e}tendue} of $1.22\,$m$^2\,$deg$^2$.  WFCAM
is likely to remain as the near--infrared imager with the largest {\em
\'{e}tendue} in the world until completion of the near--infrared
camera for VISTA (Dalton et al.  2004).  Consequently WFCAM provides
the opportunity for new wide surveys, to depths substantially deeper
than reached by 2MASS.  WFCAM is a common--user instrument, but it is
anticipated that the majority of WFCAM observing time will be devoted
to UKIDSS.

The scope, layout, and broad science goals of the five components of
UKIDSS are described by Lawrence et al.  (2006).  There are three
surveys targeting extraGalactic fields.  The Large Area Survey (LAS)
is a wide, relatively shallow survey that will cover 4000$\,$deg$^2$
from within the footprint of the SDSS, in the four bands $YJHK$.  The
depth, $K=18.4$, will be some 3$\,$mag.  deeper than 2MASS.  The Deep
Extragalactic Survey (DXS), is of intermediate depth, $K=21.0$, and
will cover 35$\,$deg$^2$ in $J$ and $K$.  The deepest survey is the
Ultra Deep Survey (UDS), which will cover 0.78$\,$deg$^2$ in $JHK$ to
a depth $K=23.0$.  Then there are two surveys targeting Galactic
fields.  The Galactic Plane Survey (GPS) will cover some
1800$\,$deg$^2$, defined by the sections of the Galactic--latitude
band $-5^{\circ}<b<+5^{\circ}$ that are contained within the
Declination limits $-15^{\circ}<\delta<+60^{\circ}$.  This region will
be imaged in $JHK$ to a depth $K=19.0$.  Finally, the Galactic
Clusters Survey will image 11 stellar open clusters and
star--formation associations, covering 1400$\,$deg$^2$, in all five
bands $ZYJHK$, to a depth $K=18.7$.

For typical observing programmes, WFCAM will accumulate data at the
rate of approximately 1Tb per week.  All WFCAM data are processed by
an automated pipeline, described by Irwin et al.  (2006).  The
pipeline flat-fields the data, subtracts the counts from the
background sky, detects and parameterises objects, and performs the
photometric and astrometric calibrations.  The reduced frames and
catalogues are ingested into the WFCAM Science Archive, described by
Hambly et al.  (2006).  A process of curation results in seamless
images and catalogues, matched across bands.  Access to the data is
through a flexible query tool, which allows SQL commands for
sophisticated searches.

The present paper characterises the photometric system defined by the
WFCAM $ZYJHK$ broadband filters.  The layout of the of the paper is as
follows.  In Section \ref{sec:photomsystem} we explain the design of
the five passbands, in Section \ref{sec:synthetic} we explain the
procedure for computing synthetic colours, and in Section
\ref{sec:ingredients} assemble the required ingredients.  In Section
\ref{sec:colours} we present the synthetic colours of a variety of
non--degenerate and degenerate stars, galaxies, and quasars.  Finally,
in Section \ref{sec:colourterms} we compute colour equations between
certain WFCAM, SDSS, and 2MASS filters.  A future paper (Hodgkin et
al. 2006, in preparation) will describe the results of calibration 
observations with
the instrument, and present the measured colour terms.  These two
papers, then, are similar in aim to the SDSS papers by Fukugita et al.
(1996) and Smith et al.  (2002).

\section{The WFCAM $Z$$Y$$J$$H$$K$ Photometric System}
\label{sec:photomsystem}

Extensive work on the specification of optimal passbands in the
near-infrared has led to the development of the Mauna Kea
Observatories (MKO) near-infrared filter set that includes
$JHKL^{\prime}M^{\prime}$ bandpasses (Simons \& Tokunaga 2002;
Tokunaga, Simons \& Vacca 2002).  The MKO filters have been adopted
widely and the filters chosen for the WFCAM $J$, $H$ and $K$ bands
have been manufactured to the MKO specifications.

Key science goals of UKIDSS include a census of very low temperature
brown dwarfs (Section \ref{sec:ydwarfs}) and the identification of
quasars with redshifts beyond the current maximum reached by SDSS,
$z=6.4$ (Section \ref{sec:quasars}).  To enable such goals, a novel
feature of UKIDSS is the extension of imaging observations to
wavelengths shortward of $1.2\,\mu$m where the availability of
broadband colours improves the discrimination between brown dwarfs and
high-redshift quasars.

At the interface between conventional optical and near-infrared
observations the SDSS $z$-band has in practice become the ``standard''
and the availability of extensive sky-coverage in the $z$-band from
the SDSS is an important element for the exploitation of the UKIDSS
LAS.  There is thus an argument for incorporating the SDSS $z$-band,
or a very close approximation to the band, into the UKIDSS filter set.
However, while the SDSS $z$-band imaging is extensive, a feature of the
passband is an extended tail in the response curve extending to long
wavelengths.  The SEDs of low-mass stars, brown dwarfs and
high-redshift quasars involve both steeply rising flux distributions
towards red wavelengths, coupled with large spectral discontinuities.
As a result, the magnitude difference between the SDSS $z$-band and a
passband with a more rectangular or Gaussian transmission profile can
be large for red objects.  Furthermore, the form of the extended red
tail in the SDSS $z$-band is defined by the declining CCD detector
quantum efficiency and is also affected by strong atmospheric
absorption at $0.93-0.97\,\mu$m, making the synthesis of a passband
closely similar to the SDSS $z$-band impractical.  For these reasons a
new $Z$-band filter has been designed for use with WFCAM.  The
specification involves: an effective wavelength of $0.882\,\mu$m
(close to the effective wavelength of the SDSS $z$-band), close to
constant transmission over a $0.06\,\mu$m wavelength range, and
cut--on and cut--off profiles very similar to the MKO $JHK$ filters.

The wavelength interval between $0.9\,\mu$m and $1.2\,\mu$m has, until
now, been largely unexploited.  The low sensitivity of most optical
CCD detectors at $\sim 1\,\mu$m, coupled with the relatively bright
sky-background curtailed interest from optical astronomers.  For
infrared astronomers, the lack of large near-infrared detectors, that
would enable wide-field surveys to be undertaken, has meant that
observations at $\sim 1\,\mu$m would be of interest in only a few
specialised applications.  For the first time, the technological
advance represented by the commissioning of WFCAM allows wide-field
observations with an instrument that possesses high throughput at
$\sim 1\,\mu$m.  Scientifically, the aim of detecting bright quasars
with redshifts as high as $z=7.2$, provides a strong motivation for
obtaining broadband observations at wavelengths between the $Z$ and
$J$ bands.

To this end, Warren \& Hewett (2002) proposed a specification for a
new $Y$-band filter with an effective wavelength of $1.03\,\mu$m.
Hillenbrand et al.  (2002) have also defined and characterised a
$Y$-band filter with a similar effective wavelength but a
substantially greater width.  The rationale for the choice of the
narrower $Y$ band involves avoiding the wavelength region that suffers
from significant atmospheric absorption by adopting a blue cut-on
wavelength of $\simeq 0.97\,\mu$m.  At the red end of the passband,
the onset of very strong emission from OH lines occurs at
$1.07\,\mu$m.  This increases the signal from the sky background
significantly and the Warren \& Hewett $Y$-band cut--off wavelength is
chosen to be $\simeq 1.07\,\mu$m \ in order to reduce the effect of
the sky background.  The narrower passband also has the significant
advantage of allowing improved discrimination between high-redshift
quasars and T dwarfs, by some 0.15$\,$mag.  The narrower $Y$-band
specification has therefore been adopted for the UKIDSS $Y$-band
filter and several other observatories have also decided to
incorporate the narrower $Y$-band specification in their filter sets
(A.  Tokunaga, private communication).

\section{Synthetic Photometry}
\label{sec:synthetic}

All calculations of magnitudes and colours have been undertaken using
the {\bf synphot} package in the Space Telescope Science Data Analysis
System.  The mean flux density, $f_{\lambda}(P)$, in a broad passband
defined by a dimensionless bandpass throughput function, $P(\lambda)$,
is calculated as:

\begin{equation}
f_{\lambda}(P) = \frac {\int P(\lambda) f_{\lambda}(\lambda) \lambda
d\lambda}{\int P(\lambda) \lambda d\lambda}
\end{equation}

where $f_{\lambda}(\lambda)$ is the flux density of the target object
(Synphot User Guide 1998\footnote{{\tt
http://www.stsci.edu/resources/software\_hardware/ stsdas/synphot}}).
The associated zero-point is calculated by evaluating the same
expression for a spectrophotometric standard star, Vega in our case,
``observed'' using the identical passband function $P(\lambda)$.  The
technique is thus differential in nature and mimics the procedure
undertaken when performing actual photometric observations, calibrated
by observations of standard stars.  The shape of the throughput
function is the key element in the calculation, although, for all but
the most pathological of target SEDs, errors in the form of the
throughput function cancel to first-order because the zero-point
defined by the standard star is calculated using the same throughput
function.

{\bf Synphot} employs a reference SED for Vega from Bohlin \& Gilliland (2004).
Since in the system of the UKIRT Faint Standards (Hawarden et al.  2001), that
will be used to calibrate UKIDSS, Vega has zero magnitude in $J$, $H$, and $K$,
whereas in the Johnson $U$$B$$V$ system Vega has non--zero magnitudes, care must
be taken in defining zero-points.  {\em In this paper the zero-points in all the
{\rm UKIDSS}, {\rm SDSS}, and {\rm 2MASS} bands are defined by Vega having zero
magnitude.}  Zero-point offsets to the {\rm AB} system used to bring the SDSS
bands onto the Vega system are provided in Section \ref{sec:ab}.  Transfer to a
system in which Vega has non-zero magnitude (as is the case for some
near--infrared systems) can be achieved by a simple zero-point shift.

\subsection{Accuracy of zero points}

{\em We assume that the spectrum of Vega provided by Bohlin \& Gilliland
(2004) is an accurate representation of the true SED.}  Employing this
spectrum as the basis for the calculations presented here ensures that
the results are reproducible by others. If an improved
determination of the Vega spectrum becomes available it would
therefore be a simple matter to compute adjustments to any quantities
provided here.

Extensive work has been undertaken to determine the absolute
flux-distribution of Vega from the ultraviolet through near-infrared
wavelengths (e.g.  Bohlin \& Gilliland (2004); Cohen, Wheaton \&
Megeath (2003)) but uncertainty at the 2\% level remains (see Tokunaga
\& Vacca (2005) for a recent discussion).  A particular concern is
whether the near-infrared absolute fluxes of Vega should be increased
by $\simeq 2\%$ relative to those in the optical.  While the basis for
the calculation of the magnitudes and colours presented here is clear,
i.e.  the Bohlin \& Gilliland Vega spectrum has zero magnitude for all
passbands, if systematic offsets in magnitude or colour at the few per
cent level are an issue for a particular application, careful reference
should be made to the above emphasised statements on the zero-points
definition, and the adopted SED for Vega.

\section{Response Functions}
\label{sec:ingredients}

Passband response functions have been computed by taken into
consideration all wavelength--dependent quantities, including
atmospheric absorption, mirror reflectivity, filter transmission, and
array quantum efficiency. Wavelength--independent quantities are
irrelevant for synthesising colours, but have been included by
normalising the computed curves to the measured total system
throughput established from observations of standard stars. This
result is preliminary, and the overall wavelength--independent
normalisation applied is subject to revision.

\subsection{Atmospheric Transmission}

The effect of atmospheric transmission was quantified using the ATRAN
code (Lord 1992).  Models were generated over the wavelength range
$0.8-2.5\,\mu$m \ with a resolution of $3\,$\AA \ using atmospheric
parameters appropriate for observations from Mauna Kea.  Using four
values for the airmass (1.0, 1.3, 1.6 and 2.0) and four values for the
precipitable water-vapour (1.0, 2.0, 3.0 and 5.0$\,$mm) resulted in a
total of 16 models.  The vast majority of the UKIDSS observations are
expected to be performed at airmass $\le 1.3$ with a water-column of
$\simeq 1.0\,$mm.  However, more extreme airmass values and
water-columns up to $5\,$mm were included to allow investigation of
the passband behaviour in both poor conditions on Mauna Kea and for
sites with much larger values of precipitable water-vapour, such as
those employed for 2MASS.  The impact of Rayleigh scattering and
aerosol scattering (Hayes \& Latham 1975) on the shape of the
passbands is insignificant given our target photometric accuracy and
no attempt has been made to include their effects.

\begin{table}
\caption{Wavelength dependence of array detector q.e.}
\label{qe}
\centering
\begin{scriptsize}
\begin{tabular}{ccllllllll}\\ \hline
$\lambda$ $\mu$m & 0.8  & 0.9  & 1.0  & 1.1  & 1.2  & 1.4  & 1.6  &
2.2  \\ 
q.e.  & 0.53 & 0.55 & 0.54 & 0.54 & 0.54 & 0.57 & 0.62 & 0.61 \\  \hline
\end{tabular}
\end{scriptsize}
\end{table}

\subsection{Telescope, Instrument and Detector}

Light entering the telescope undergoes three reflections from
aluminium-coated mirrors (telescope primary, telescope secondary and
one reflection within WFCAM) before reaching the detectors.  The
variation in reflectivity at short infrared wavelengths, relevant for
the $Z$ and $Y$ bands, is significant and three reflections from
aluminium have been incorporated in the calculation of the passbands
using data from Hass \& Hadley (1963).

Transmission measurements over the wavelength range of interest for
the remaining optical components within the WFCAM instrument show no
variations exceeding 1\%, and constant transmission, independent of
wavelength, has therefore been adopted.

The quantum efficiency of the Rockwell Hawaii-II $2048 \times 2048$ HgCdTe
detectors increases approximately linearly with wavelength.  The change across
the individual passbands amounts to only a few per cent but the variation in
quantum efficiency has been incorporated in the calculation of the passbands
using data for a typical device (Table \ref{qe}).  The measurements at 1.2, 1.6,
and 2.2$\mu$m are the values supplied by the manufacturer, multiplied by 0.8.
This factor accounts for an error in the computed gain used by the manufacturer,
as a consequence of covariance between pixels.  To obtain the remaining values
in the table, the relative quantum efficiency across the interesting wavelength
range was measured by the WFCAM team (Casali et al, 2006), and the values were
normalised to the manufacturer's points.  We have interpolated between the
points in the table, with the quantum efficiency assumed to be constant for
wavelengths $>2.2\,\mu$m.

\begin{figure*}
\begin{center}
\includegraphics[width=18cm]{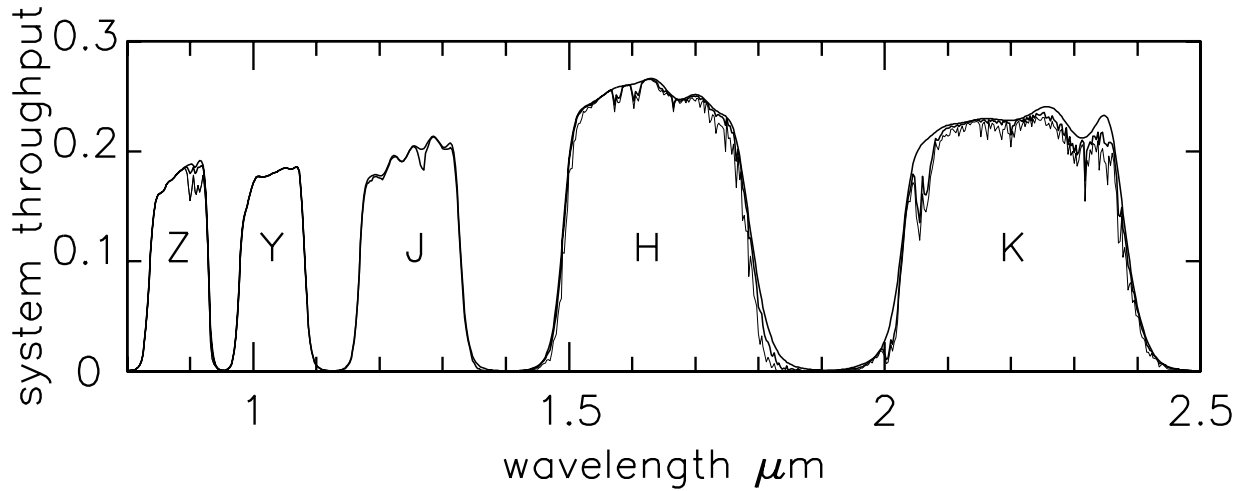}
\caption{\label{filters}Transmission curves, from above atmosphere to
    detector, for the WFCAM filter set. Three curves are shown for each
    band. The upper, thick, curve is for no atmospheric
    absorption. The middle, thick, curve is for the default
    atmospheric conditions (1.3 airmass, 1.0mm water), and the lower, 
thin,
    curve is for extreme atmospheric conditions (2.0 airmass, 5.0 mm
    water). These curves have been computed on the basis of measurement
    or modeling of the relevant wavelength--dependent quantities. An
    overall wavelength--independent normalisation has been applied
    to match the photon count rate measured for standard stars with the
    instrument. The calculation is preliminary and the
    wavelength--independent normalisation is subject to revision.
}
\end{center}
\end{figure*}

\subsection{Filters}

The filter transmission curves are from laboratory measurements at
$10\,$\AA \ intervals, made by the WFCAM team. The measurements were
undertaken at room temperature.  The dependence of the filter cut-on
and cut-off wavelengths as a function of temperature was measured by
the manufacturers.  Filter transmission curves appropriate for the
operating temperature ($120\,$K) of the WFCAM instrument were then
derived by us by applying the measured temperature dependence at the
cut--on and cut--off wavelengths, with linear interpolation adopted for
wavelengths between the two reference points.  For the $Z$ filter the
measured temperature dependence was consistent with no change and the
transmission profile measured at room temperature was used. At the
other extreme, the $K$-filter exhibits a shift of $\simeq
0.035\,\mu$m, or $\sim 10\%$ of the bandwidth, and the use of the
temperature--corrected profiles is important when calculating
passbands.

The transmission measurements were undertaken from $0.4\,\mu$m \ out
to beyond $3.0\,\mu$m, where the WFCAM detectors possess no
sensitivity.  With the exception of an insignificant narrow ``blue
leak'' of height $1\%$ at $\simeq 0.792\,\mu$m \ for the $Y$ filter,
no leaks above the $0.1\%$ specification requirement were present in
any of the filter transmission curves.

The transmission curves presented here are derived from a single
filter in each band --- four filters per band are necessary to cover
the detector arrays in WFCAM --- but intercomparison of the four
transmission curves in each band shows that the variations between
filters produce differences in the photometry of at most $0.01$
magnitudes.

\subsection{Summary: WFCAM system throughput curves}

  Fig. 1 shows the $ZYJHK$ passband transmission curves for: i) no
atmosphere, ii) typical atmospheric conditions, airmass 1.3 and
$1.0\,$mm of precipitable water-vapour, and iii) extreme atmospheric
conditions, airmass 2.0 and $5.0\,$mm of precipitable water-vapour.

The MKO $JHK$ filter transmissions were developed with the aim of
minimising the effects of variable atmospheric transmission due
primarily to the change in the water content of the atmosphere.  The
same philosophy was adopted in the design of the new $Z$ and $Y$
filters.  The result is a reassuring stability of the UKIDSS $ZYJHK$
system over a wide range of atmospheric conditions.  There are
essentially no differences in the photometry at the $>0.01\,$mag level
for objects with non-pathological SEDs for observations made through
precipitable water-vapour columns of $\le 3.0\,$mm and at airmasses in
the range $1.0-1.6$.

The insensitivity of the WFCAM $ZYJHK$ system to atmospheric properties is
evident from consideration of observations of the full range of stellar types
included in the Bruzual--Persson--Gunn--Stryker atlas (Section 5.1.1).  Even for
observations made at an airmass of 2.0 through an extreme atmospheric model for
Mauna Kea that includes $5.0\,$mm of water-vapour (Fig.  1), only in the
$Z$-band, for a narrow range of stellar type (M6-M8), are differences in the
photometry that exceed $>0.01\,$mag level evident.  The differences in the
$Z$-band, compared to observations made in excellent conditions, amount to no
more than $0.015\,$ mag for the narrow range of late-M spectral types.
Similarly, for the adopted quasar SED (Section 5.4), differences between
observations made through the extrema of the atmospheric models differ by less
than $0.02\,$mag for redshifts $z < 6.2$.  Only at redshifts ($z \sim 6.5$)
where the very strong spectral discontinuity at Lyman-$\alpha$ $1216\,$\AA \
coincides with the atmospheric absorption at $9000-9200\,$\AA \ in the $Z$-band
does the photometry differ by more than $0.1\,$mag, reaching a maximum of
$0.2\,$mag.

Synthetic photometry for such truly unusual SEDs observed through the
most extreme atmospheric models can be obtained using the appropriate
transmission curves.  However, for the calculations presented in this
paper we have adopted the passband transmission curves based on
observations at an airmass of 1.3 through an atmosphere with $1.0\,$mm
of precipitable water-vapour, typical of the conditions under which
the majority of UKIDSS data are likely to be obtained. These default
$ZYJHK$ passband transmissions are provided in Tables 2--6, each of which
lists in col. 1 the wavelength in $\mu$m and in col. 2. the fractional
throughput.

\subsection{SDSS and 2MASS Passbands}

The SDSS passbands are based on data obtained by J.  Gunn to calculate the
``June 2001'' version of the SDSS passbands, which were kindly supplied by X.
Fan.  The passbands are identical to those provided on-line with the recent SDSS
DR4 release.  The passbands are appropriate for the observation of point sources
from Apache Point Observatory at an airmass of 1.3.

The 2MASS passbands were derived by taking the relative spectral
response curves given by Cohen et al. (2003), dividing by wavelength
and renormalising to produce a transmission function, $P(\lambda)$, as
employed in equation 1.

\setcounter{table}{6}

\begin{table}
\caption{Conversion to AB mag.}
\label{aboffsets}
\centering
\begin{scriptsize}
\begin{tabular}{ccrlr}\\ \hline
band & $\lambda_{\mathrm{eff}}$ & \multicolumn{1}{c}{S}  & flux density & 
 \multicolumn{1}{c}{AB} \\
     &   $\mu$m        & \multicolumn{1}{c}{Jy} & 
 W\,m$^{-2}\mu^{-1}$ & \multicolumn{1}{c}{offset} \\ \hline
$u$ &  0.3546 & 1545 &	$3.66\times10^{-8}$  &  0.927 \\
$g$ &  0.4670 & 3991 & 	$5.41\times10^{-8}$  & -0.103 \\
$r$ &  0.6156 & 3174 &	$2.50\times10^{-8}$  &  0.146 \\
$i$ &  0.7471 & 2593 &	$1.39\times10^{-8}$  &  0.366 \\
$z$ &  0.8918 & 2222 &	$8.32\times10^{-9}$  &  0.533 \\
$Z$ &  0.8817 & 2232 &	$8.59\times10^{-9}$  &  0.528 \\
$Y$ &  1.0305 & 2026 &	$5.71\times10^{-9}$  &  0.634 \\
$J$ &  1.2483 & 1530 &	$2.94\times10^{-9}$  &  0.938 \\
$H$ &  1.6313 & 1019 &	$1.14\times10^{-9}$  &  1.379 \\
$K$ &  2.2010 &  631 &	$3.89\times10^{-10}$ &  1.900 \\ \hline
\end{tabular}
\end{scriptsize}
\end{table}

\subsection{AB system offsets}
\label{sec:ab}

In Table \ref{aboffsets} we list the filters and provide in col. 2 the
computed effective wavelength, defined according to Schneider, Gunn \&
Hoessel (1983) (see equation 3 in Fukugita et al.  1996). Cols 3 and 4
list the flux-density of an object spectrum, of constant flux density
in both Jy and ${\rm W}{\rm m}^{-2} \mu^{-1}$, over each passband,
that corresponds to zero magnitude. The former quantity is quantified
in another way in col. 5, as the magnitude offset to convert the
Vega-based magnitudes onto the AB-magnitude system (Oke and Gunn,
1983), where we have used the definition ${\rm AB}_{\nu} = -2.5\, {\rm
log}\, f_{\nu}\, ({\rm erg\,s^{-1} cm^{-2} Hz^{-1}}) + 48.60$. The
values for the $J$, $H$, and $K$ bands computed by Tokunaga and Vacca
(2005), for the MKO system, are virtually identical. This is a
coincidence, however, as they employed a slightly different SED for
Vega, and a different zero--point of the AB system, these two
differences closely cancelling each other.

\section{Synthetic colours of stars, galaxies, and quasars}
\label{sec:colours}

\begin{figure*}
\begin{center}
\includegraphics[width=18cm]{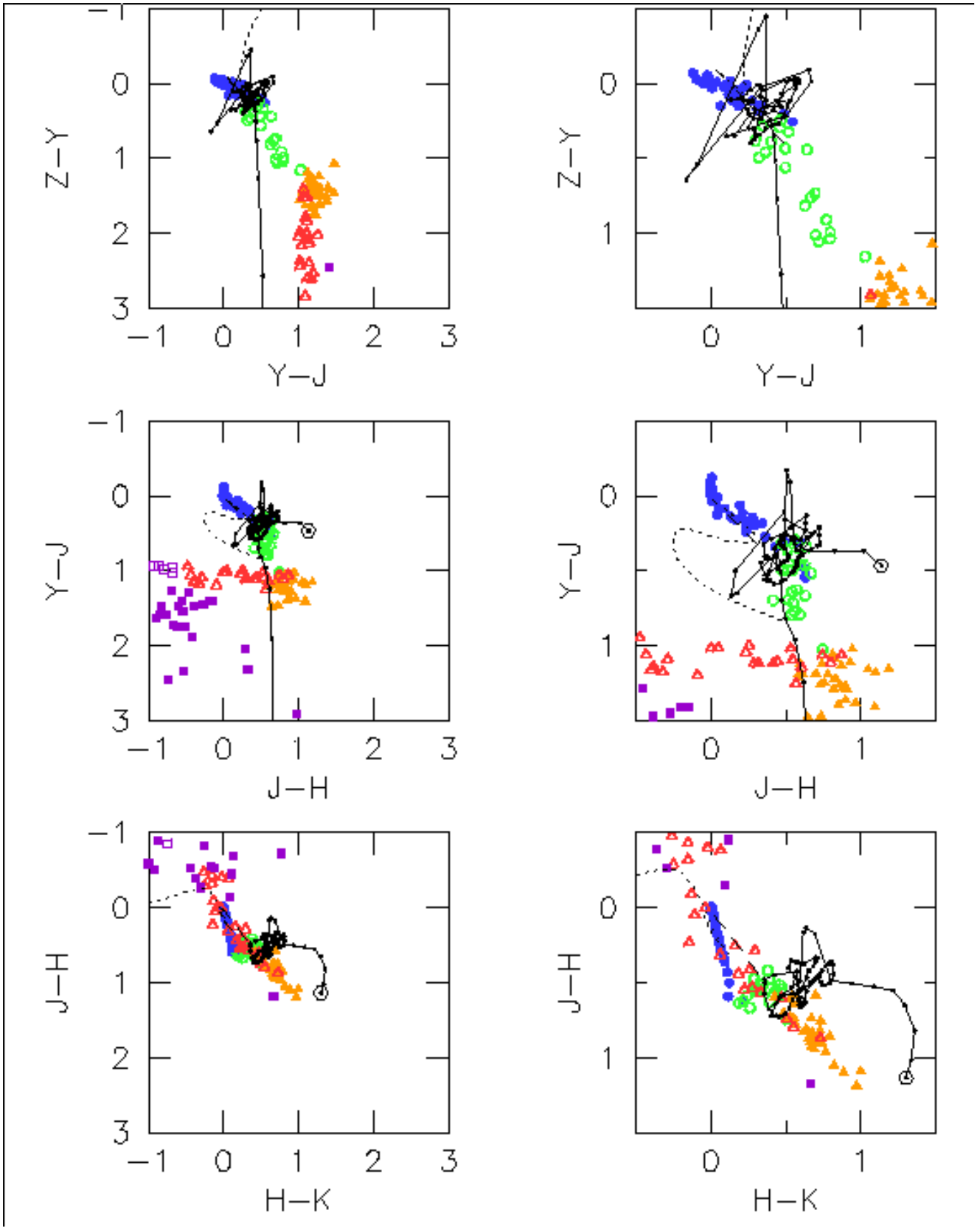}
\caption{\label{fig2colourstars}Top to bottom, three 2--colour
     diagrams of stellar sources, cycling through the colour sequence
     $Z-Y$, $Y-J$, $J-H$, $H-K$. Each right--hand panel is a $2\times$
     expanded view of part of the corresponding left--hand panel. Key:
     BPGS O--K dwarfs blue {\Large \textbullet}; M dwarfs green $
     \bigcirc$; L dwarfs orange $\blacktriangle$; T dwarfs red
     $\triangle$; Burrows model cool brown dwarfs purple
     $\blacksquare$; Marley model cool brown dwarfs purple $\square$;
     quasars $0<z<8.5$, $\Delta z=0.1$, solid black line, $z=0$ marked
     by open hexagon; H white dwarfs dotted black line, He white
     dwarfs dashed black line.}
\end{center}
\end{figure*}

\begin{figure*}
\begin{center}
\includegraphics[width=18cm]{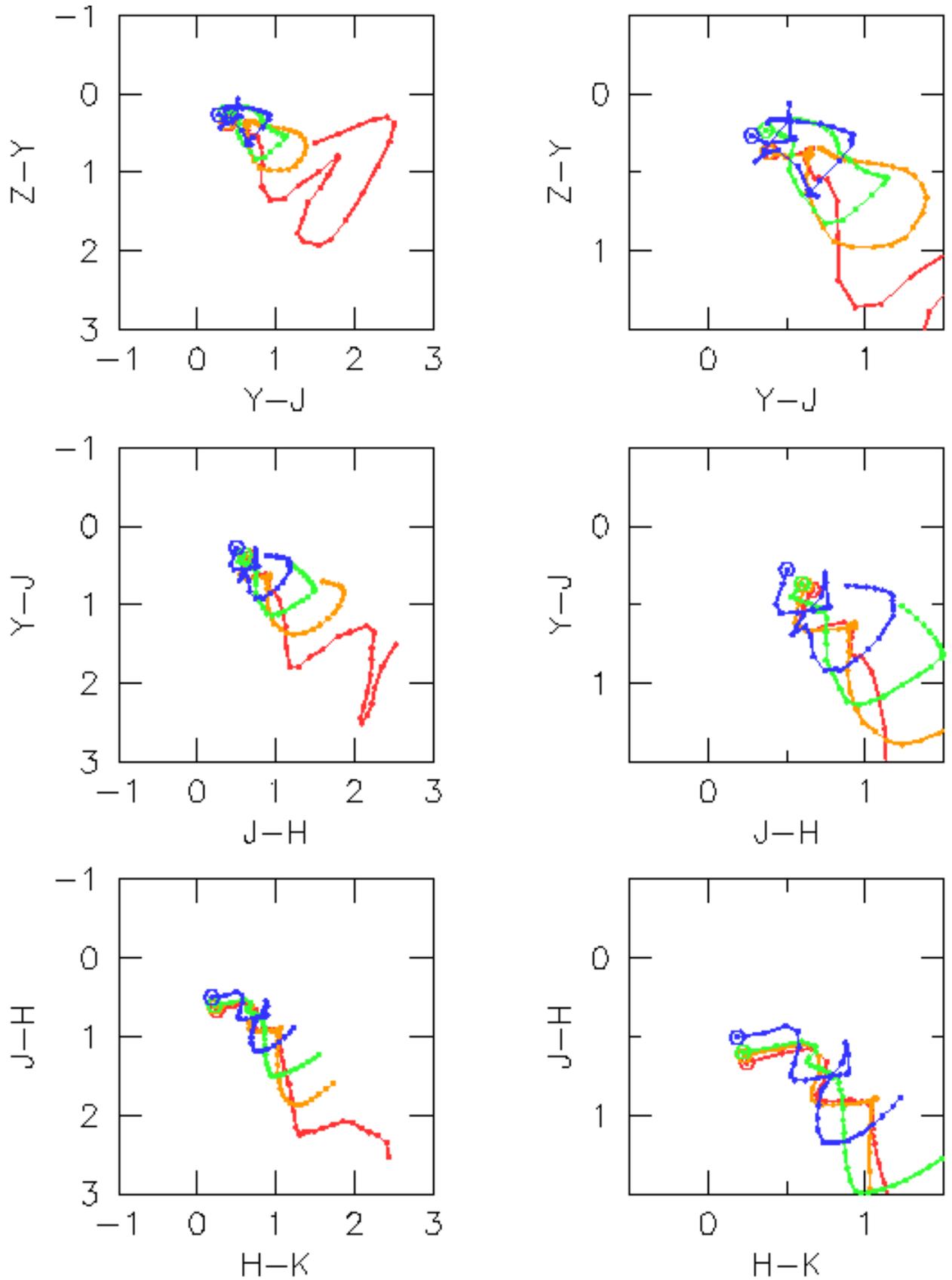}
\caption{\label{fig3colourgalaxies}Top to bottom, three 2--colour
diagrams
     of {\em hypeprz} galaxy templates, redshift range $0<z<3.6$,
     $\Delta z=0.1$, cycling through the colour sequence $Z-Y$, $Y-J$,
     $J-H$, $H-K$. Each right--hand panel is a $2\times$ expanded view of
     part of the corresponding left--hand panel. Key: E red, Sbc orange,
     Scd green, Im blue. For all curves $z=0$ is marked by an open hexagon.}
\end{center}
\end{figure*}

In this section we use the apparatus described in Sections
\ref{sec:synthetic} and \ref{sec:ingredients}, to synthesise colours
of a wide range of sources --- stars, galaxies, and quasars --- to
illustrate the colours of the types of objects expected to appear in
the surveys.  The following sub--sections detail the origins of the
spectra, and tabulate the results for each class of object.  Fig.
\ref{fig2colourstars} (for brown dwarfs, stars and quasars) and Fig.
\ref{fig3colourgalaxies} (for galaxies) illustrate the $Z$, $Y$, $J$,
$H$, $K$, colours which are discussed in the relevant sub-sections
that follow.  Because of the wide range of computed colours, in each
of the two figures the left--hand set of panels and the right-hand set
of panels show the same two--colour diagrams at different scales.  As
discussed in Section 4.4 all calculations presented employ the
passband transmission functions appropriate to observations at airmass
1.3 through an atmosphere with $1.0\,$mm of precipitable water-vapour.

The results shown here may be reproduced by taking the tabulated
passband transmissions, $P(\lambda)$ (Tables 2-6),
and evaluating the quantity
$f_\lambda(P)$ defined in equation 1, for a particular object SED
(provided in the form $f_\lambda(\lambda)$).  The zero-point is
derived by performing the same calculation for the Bohlin \& Gilliland
(2004) Vega spectrum.  Colours in the same system may be computed for
any object of specified SED by following this procedure.

\subsection{Non-degenerate stars: the BPGS spectrophotometric atlas,
and additional M dwarfs}

\begin{table*}
\caption{Colours of selected stars from the BPGS atlas}
\label{bpgs}
\centering
\begin{scriptsize}
\begin{tabular}{rrrrrrrrrrrrcrl}\\ \hline
\multicolumn{1}{c}{\ug} & \multicolumn{1}{c}{\gr} &
\multicolumn{1}{c}{\ri} & \multicolumn{1}{c}{\iz} & 
\multicolumn{1}{c}{\Zz} & \multicolumn{1}{c}{\ZY} & 
\multicolumn{1}{c}{\YJ} & \multicolumn{1}{c}{\JH} & 
\multicolumn{1}{c}{\HK} & \multicolumn{1}{c}{\JJ} & 
\multicolumn{1}{c}{\HH} & \multicolumn{1}{c}{\KK} & 
\multicolumn{1}{c}{Class} & \multicolumn{1}{c}{BPGS} & 
\multicolumn{1}{c}{Name} \\ 
 & & & & & & & & & & & & & \multicolumn{1}{c}{no.} & \\ \hline
 -0.456 & -0.066 & -0.029 &  0.031 & -0.012 & -0.069 & -0.120 &  0.006 & -0.002 &  0.007 &  0.000 & -0.001 & B9V     &   13 & HD189689       \\
 -0.043 & -0.010 &  0.011 &  0.034 & -0.010 & -0.027 & -0.094 &  0.000 & -0.008 &  0.009 &  0.002 & -0.002 & A0V     &   14 & THETA-VIR      \\
 -0.133 & -0.019 &  0.027 &  0.040 & -0.002 & -0.023 & -0.084 & -0.003 & -0.007 &  0.009 &  0.001 & -0.002 & B9V     &   15 & NU-CAP         \\
 -0.012 & -0.002 &  0.005 &  0.000 & -0.008 & -0.056 & -0.021 &  0.003 & -0.007 &  0.003 &  0.001 & -0.002 & A2V     &   16 & HR6169         \\
 -0.068 & -0.002 &  0.017 &  0.036 & -0.002 & -0.022 & -0.010 & -0.008 &  0.001 &  0.006 &  0.001 & -0.002 & A1V     &   17 & HD190849A      \\ \hline    
\end{tabular}
\begin{minipage}{165mm}
Note: The full table is published in the electronic version of the
paper. A portion is shown here for guidance regarding its form and content.
\end{minipage}
\end{scriptsize}
\end{table*}

\subsubsection{BPGS spectrophotometric atlas}

The Bruzual--Persson--Gunn--Stryker (BPGS) atlas is a database of
spectrophotometry of 175 stars, of a wide range of spectral type and
luminosity class, spanning the optical and near--ir wavelength range.
The atlas itself is unpublished, but is available at e.g.  {\tt
http://www.stsci.edu/hst/observatory/cdbs/bpgs.html}.  The optical
spectra comprise the Gunn--Stryker atlas (Gunn \& Stryker, 1983).
Most of the near-infrared data come from Strecker, Erickson \&
Witteborn (1979), while the remainder are unpublished.  Stars 1--12,
70--77, 173, and 174 have been omitted from the calculations due to
incomplete wavelength coverage.  Stars 169, and 170 were also excluded
because of an apparent mismatch between the amplitudes of the optical
and near-infrared spectra.

The synthesised colours of the BPGS stars are provided in Table
\ref{bpgs}.  The first 12 columns list colours from $u$ to $K$, and
columns $13-15$ list successively the spectral classification, the
BPGS reference number and the name of the star, taken from Gunn \&
Stryker (1983), if provided.

\begin{table*}
\caption{Colour of selected additional M dwarfs}
\label{leggettm}
\centering
\begin{scriptsize}
\begin{tabular}{rrrrrrrrrrll}\\ \hline
\multicolumn{1}{c}{\iz} & \multicolumn{1}{c}{\Zz} & 
\multicolumn{1}{c}{\ZY} & \multicolumn{1}{c}{\YJ} & 
\multicolumn{1}{c}{\JH} & \multicolumn{1}{c}{\HK} & 
\multicolumn{1}{c}{\JJ} & \multicolumn{1}{c}{\HH} & 
\multicolumn{1}{c}{\KK} & \multicolumn{1}{c}{M$_K$} &
\multicolumn{1}{c}{Class} & \multicolumn{1}{c}{Name} \\ 
\hline
  0.576 &  0.021 &  0.323 &  0.519 &  0.668 &  0.260 & -0.034 &  0.020 & -0.008 & 5.3$^{\rm a}$  & M1   & Gl 229A \\
    ... & -0.014 &  0.292 &  0.458 &  0.593 &  0.237 & -0.040 &  0.020 & -0.038 & 8.2$^{\rm b}$  & M3   & LHS 5327 \\
  0.973 &  0.032 &  0.432 &  0.498 &  0.543 &  0.303 & -0.037 &  0.026 & -0.019 & 7.8$^{\rm b}$  & M3.5 & GJ 1001A \\
  0.966 &  0.017 &  0.440 &  0.640 &  0.541 &  0.337 & -0.053 &  0.026 & -0.025 & 8.1$^{\rm a}$  & M3.5 & Gl 15B \\
  1.147 &  0.023 &  0.432 &  0.492 &  0.514 &  0.316 & -0.044 &  0.021 & -0.052 & 8.2$^{\rm a}$  & M4   & Gl 699 \\
  0.957 &  0.055 &  0.559 &  0.496 &  0.476 &  0.288 & -0.040 &  0.028 & -0.029 & 6.9 $^{\rm b}$ & M4.5 & Gl 630.1A \\
  0.858 &  0.078 &  0.815 &  0.628 &  0.571 &  0.421 & -0.040 &  0.027 & -0.032 & 9.0$^{\rm b}$  & M5.5 & GJ 1002 \\
  1.508 &  0.113 &  1.056 &  0.718 &  0.548 &  0.433 & -0.049 &  0.034 & -0.031 & 9.5$^{\rm b}$  & M5.5 & GJ 4073 \\
  1.337 &  0.077 &  0.733 &  0.694 &  0.516 &  0.457 & -0.048 &  0.031 & -0.035 & 8.5$^{\rm b}$  & M5.5 & GJ 1245A \\
  1.170 &  0.060 &  0.761 &  0.667 &  0.627 &  0.391 & -0.052 &  0.025 & -0.024 & 8.4$^{\rm b}$  & M5.5 & Gl 905 \\
  1.550 &  0.089 &  0.909 &  0.771 &  0.559 &  0.439 & -0.050 &  0.029 & -0.024 & 9.1$^{\rm b}$  & M6   & Gl 406 \\
    ... &  0.037 &  1.012 &  0.698 &  0.417 &  0.380 & -0.064 &  0.029 & -0.053 & 8.7$^{\rm b}$  & M6   & LHS 5328 \\
  1.720 &  0.120 &  1.036 &  0.794 &  0.593 &  0.455 & -0.055 &  0.034 & -0.025 & 9.7$^{\rm b}$  & M6.5 & GJ 3855 \\
  1.559 &  0.092 &  0.992 &  0.796 &  0.538 &  0.472 & -0.065 &  0.037 & -0.026 & 9.7$^{\rm b}$  & M7   & VB 8 \\
  2.299 &  0.157 &  1.154 &  1.028 &  0.747 &  0.511 & -0.051 &  0.034 & -0.016 & ...  & M8.5 & SDSS J225529.09-003433.4 \\ \hline
\end{tabular}
\begin{minipage}{165mm}
a ESA (1997) 

b van Altena, Lee \& Hoffleit (1995)
\end{minipage}
\end{scriptsize}
\end{table*}

\subsubsection{Additional M dwarfs}
\label{sec:mstars}

The BPGS atlas contains spectra of only six M dwarfs.  Therefore we
have supplemented the list with spectra of 15 M dwarfs covering the
wavelength range $i$ to $K$, collated from the following references:
Geballe et al.  (2002), Hawley, Gizis \& Reid (1996), Henry,
Kirkpatrick \& Simons (1996), Kirkpatrick, Henry \& Simons (1995),
Leggett et al.  (2000a; 2001; 2002a), McLean et al.  (2003), and Reid,
Hawley \& Gizis (1995).  In Table \ref{leggettm}, cols $1-9$ provide
colours over the range $i$ to $K$, col.  10 gives the $K$-band
absolute magnitude, M$_K$, with the source of the parallax
determination indicated, col.  11 provides the spectral classification
and col.  12 the object name.  In a few cases the spectrum does not
cover the entire $i$ band, and the $i-z$ entry is consequently blank.
The spectra were observed in a small number (e.g.  three) of sections,
each spectrophotometrically calibrated.  To account for differential
slit losses, the sections of spectra were calibrated absolutely using
published photometry, in $J$, $H$, and $K$, and frequently in a wide
$Z$ filter that covers the entire range of the WFCAM $Z$ and $Y$
filters.  The result is that our synthesised colours to a large extent
simply reproduce the published colours in the $J$, $H$, and $K$ bands,
whereas colours that include the $i$, $Z$, and $Y$ bands involve a
degree of interpolation or extrapolation, and therefore will be less
accurate.  The uncertainties in the spectral fluxes (at wavelengths
within the WFCAM photometric passbands) are typically 3\%.  Deviations
between the measured $JHK$ colours and the slopes of the spectra are
1-5\%.  Hence the errors in the predicted colours of the M-dwarfs, and
the L and T dwarfs discussed in Section 5.2.1, are expected to be
0.03-0.06\,mag.

In Fig.  \ref{fig2colourstars} the colours of BPGS dwarfs covering the range of
spectral types from O to K, are plotted as blue filled circles.  M dwarfs are
plotted as green open circles, and extend the colour sequence displayed by 
the O to K dwarfs to redder colours.

\subsection{Degenerate stars: L and T dwarfs, model Y dwarfs, and cool
white dwarfs}


\begin{table*}
\caption{Colours of selected L and T dwarfs}
\label{landt}
\centering
\begin{scriptsize}
\begin{tabular}{rrrrrrrrrlll}\\ \hline
\multicolumn{1}{c}{\iz} & \multicolumn{1}{c}{\Zz} & 
\multicolumn{1}{c}{\ZY} & \multicolumn{1}{c}{\YJ} & 
\multicolumn{1}{c}{\JH} & \multicolumn{1}{c}{\HK} & 
\multicolumn{1}{c}{\JJ} & \multicolumn{1}{c}{\HH} & 
\multicolumn{1}{c}{\KK} & \multicolumn{1}{c}{M$_K$} & 
\multicolumn{1}{c}{Class} & \multicolumn{1}{c}{Name} \\ 
\hline
  2.077 &  0.100 &  1.069 &  1.476 &  0.643 &  0.549 & -0.085 &  0.041 & -0.016 & 10.5$^{\rm a}$ & L1 & 2MASS J03454316+2540233 \\
  1.996 &  0.188 &  1.322 &  1.193 &  0.610 &  0.527 & -0.073 &  0.045 &  0.007 & 10.0$^{\rm a}$ & L1 & 2MASS J07464256+2000321AB \\
  2.165 &  0.232 &  1.419 &  1.412 &  1.092 &  0.997 & -0.096 &  0.058 & -0.020 & ... & L3 & 2MASS J00283943+1501418 \\
    ... &  0.251 &  1.460 &  1.289 &  0.819 &  0.729 & -0.094 &  0.056 & -0.024 & ... & L3 & SDSS J111320.16+343057.9 \\
  2.064 &  0.187 &  1.385 &  1.397 &  0.850 &  0.718 & -0.088 &  0.055 & -0.020 & 11.3$^{\rm a}$ & L3 & DENIS-P J1058.7-1548 \\
  2.137 &  0.248 &  1.503 &  1.366 &  0.896 &  0.665 & -0.085 &  0.048 & -0.011 & 11.5$^{\rm b}$ & L3 & GD 165B \\
  2.138 &  0.215 &  1.458 &  1.472 &  0.733 &  0.550 & -0.100 &  0.055 & -0.004 & 11.3$^{\rm a}$ & L4 & 2MASS J00361617+1821104 \\
    ... &  0.168 &  1.235 &  1.274 &  0.899 &  0.728 & -0.102 &  0.057 & -0.017 & ... & L4.5 & SDSS J085116.20+181730.0 \\
  2.337 &  0.254 &  1.418 &  1.263 &  0.857 &  0.792 & -0.093 &  0.056 & -0.015 & 11.4$^{\rm b}$ & L4.5 & LHS 102 B \\
  2.139 &  0.231 &  1.392 &  1.291 &  0.902 &  0.738 & -0.096 &  0.058 & -0.017 & ... & L4.5 & SDSS J083506.16+195304.4 \\
  2.719 &  0.214 &  1.359 &  1.230 &  0.821 &  0.643 & -0.096 &  0.054 & -0.013 & 11.8$^{\rm c}$ & L5   & SDSS J053951.99-005902.0 \\
  2.789 &  0.196 &  1.443 &  1.062 &  0.810 &  0.636 & -0.102 &  0.056 &  0.004 & 11.4$^{\rm a}$ & L5.5 & DENIS-P J0205.4-1159AB \\
    ... &  0.317 &  1.577 &  1.383 &  0.963 &  0.764 & -0.101 &  0.056 & -0.017 & ... & L5.5 & SDSS J134203.11+134022.2\\
  2.884 &  0.181 &  1.411 &  1.158 &  1.186 &  0.971 & -0.105 &  0.057 & -0.029 & 12.6$^{\rm c}$ & L5.5 & SDSS J010752.33+004156.1 \\
  1.472 &  0.194 &  1.375 &  1.195 &  0.778 &  0.563 & -0.111 &  0.052 & -0.009 & ... & L5.5 & SDSS J020608.97+223559.2 \\
  2.949 &  0.196 &  1.616 &  1.182 &  1.097 &  0.884 & -0.094 &  0.056 & -0.024 & 12.7$^{\rm ac}$ & L6 & 2MASS J08251968+2115521 \\
    ... &  0.230 &  1.419 &  1.261 &  0.730 &  0.525 & -0.107 &  0.052 & -0.011 & ... & L6 & SDSS J103321.92+400549.5 \\
    ... &  0.188 &  1.286 &  1.126 &  0.738 &  0.669 & -0.105 &  0.062 & -0.011 & ... & L6 & SDSS J162255.27+115924.1  \\
  2.494 &  0.257 &  1.615 &  1.284 &  0.885 &  0.695 & -0.103 &  0.057 & -0.011 & 11.1$^{\rm a}$ & L6 & DENIS-P J1228.2-1547AB \\
  1.317 &  0.298 &  1.501 &  1.246 &  0.756 &  0.732 & -0.086 &  0.048 & -0.028 & ... & L6 & SDSS J065405.63+652805.4 \\
    ... &  0.390 &  1.747 &  1.229 &  0.834 &  0.675 & -0.107 &  0.054 & -0.013 & ... & L6.5 & SDSS J141659.78+500626.4 \\
    ... &  0.202 &  1.280 &  1.188 &  0.699 &  0.484 & -0.100 &  0.053 & -0.013 & ... & L6.5 & SDSS J142227.25 221557.1 \\
  3.105 &  0.168 &  1.461 &  1.160 &  1.049 &  0.822 & -0.108 &  0.056 & -0.009 & 12.9$^{\rm a}$ & L7.5 & 2MASS J16322911+1904407 \\
    ... &  0.173 &  1.184 &  1.125 &  0.594 &  0.425 & -0.113 &  0.058 & -0.021 & ... & L7.5 & SDSS J112118.57+433246.5 \\
    ... &  0.176 &  1.406 &  1.112 &  0.928 &  0.667 & -0.112 &  0.057 &  0.004 & 12.3$^{\rm c}$ & L8 & SDSS J003259.36+141036.6 \\
    ... &  0.247 &  1.631 &  1.089 &  0.827 &  0.618 & -0.124 &  0.060 & -0.004 & ... & L9 & SDSS J100711.74+193056.2 \\
  3.089 &  0.198 &  1.470 &  1.133 &  0.870 &  0.653 & -0.107 &  0.055 & -0.013 & ... & L9 & 2MASS J09083803+5032088 \\
  3.201 &  0.197 &  1.542 &  1.025 &  0.942 &  0.688 & -0.117 &  0.060 &  0.034 & ... & L9 & 2MASS J03105986+1648155 \\
  1.765 &  0.354 &  1.738 &  1.215 &  0.587 &  0.546 & -0.112 &  0.052 & -0.008 & ... & L9.5 & SDSS J080531.80+481233.0 \\
    ... &  0.272 &  1.664 &  1.193 &  0.583 &  0.698 & -0.143 &  0.050 & -0.001 & ... & L9.5 & SDSS J151114.66+060742.9 \\
  2.605 &  0.184 &  1.513 &  1.066 &  0.870 &  0.730 & -0.135 &  0.062 & -0.004 & ... & T0 & SDSS J085834.64+325629.1 \\
  3.044 &  0.220 &  1.517 &  1.114 &  0.799 &  0.550 & -0.125 &  0.058 & -0.003 & 12.0$^{\rm c}$ & T0 & SDSS J042348.57-041403.5 \\
    ... &  0.148 &  1.410 &  1.066 &  0.743 &  0.506 & -0.140 &  0.058 &  0.001 & ... & T0 & SDSS J120747.17+024424.8 \\
  1.527 &  0.422 &  2.133 &  1.148 &  0.604 &  0.472 & -0.153 &  0.057 &  0.023 & ... & T1 & SDSS J103931.35+325625.5 \\
  1.903 &  0.276 &  1.787 &  1.097 &  0.546 &  0.224 & -0.147 &  0.059 &  0.045 & ... & T1.5 & SDSS J090900.73+652527.2 \\
  3.684 &  0.310 &  2.020 &  1.257 &  0.567 &  0.332 & -0.179 &  0.067 &  0.041 & ... & T2   & SDSS J075840.33+324723.4  \\
  4.712 &  0.325 &  1.989 &  1.021 &  0.534 &  0.275 & -0.168 &  0.063 &  0.054 & 13.1$^{\rm cd}$ & T2 & SDSS J125453.90-012247.4 \\
  1.550 &  0.223 &  1.835 &  1.121 &  0.414 &  0.258 & -0.170 &  0.053 &  0.058 & ... & T2.5 & SDSS J143945.86+304220.6 \\
    ... &  0.280 &  2.048 &  1.004 &  0.255 &  0.160 & -0.151 &  0.056 &  0.064 & ... & T3 & 2MASS J12095613-1004008 \\
    ... &  0.313 &  2.100 &  1.119 &  0.287 &  0.295 & -0.193 &  0.062 &  0.046 & ... & T3 & SDSS J153417.05+161546.1 \\
  3.352 &  0.296 &  1.991 &  1.105 &  0.443 &  0.183 & -0.187 &  0.057 &  0.056 & 12.8$^{\rm cd}$ & T3 & SDSS J102109.69-030420.1 \\
    ... &  0.390 &  2.157 &  1.044 &  0.234 & -0.144 & -0.179 &  0.045 &  0.076 & ... & T3 & SDSS 120602.51+281328.7 \\
  1.976 &  0.509 &  2.608 &  1.122 &  0.317 &  0.062 & -0.212 &  0.070 &  0.090 & ... & T3.5 & SDSS J121440.94+631643.4 \\
    ... &  0.487 &  2.447 &  1.020 &  0.050 & -0.108 & -0.188 &  0.049 &  0.107 & ... & T4 & 2MASS J22541892+3123498 \\
  4.590 &  0.491 &  2.526 &  1.198 & -0.090 & -0.131 & -0.211 &  0.046 &  0.119 & 13.7$^{\rm acd}$ & T4.5 & 2MASS J05591914-1404488 \\
  3.357 &  0.423 &  2.372 &  1.021 &  0.001 & -0.038 & -0.203 &  0.048 &  0.104 & ... & T4.5 & SDSS J092615.38+584720.9 \\
    ... &  0.612 &  2.836 &  1.091 & -0.288 & -0.251 & -0.222 &  0.041 &  0.110 & ... & T4.5 & SDSS J135852.68+374711.9 \\
  4.009 &  0.477 &  2.614 &  1.178 & -0.321 & -0.152 & -0.258 &  0.038 &  0.131 & 15.4$^{\rm acd}$ & T6 & SDSS J162414.37+002915.6 \\
  4.120 &  0.307 &  2.397 &  1.141 & -0.382 &  0.066 & -0.262 &  0.036 &  0.138 & 15.1$^{\rm cd}$ & T6.5 & SDSS J134646.45-003150.4 \\
  4.291 &  0.567 &  3.006 &  1.164 & -0.399 & -0.023 & -0.254 &  0.016 &  0.103 & 15.6$^{\rm e}$ & T7 & Gl 229B \\
  4.545 &  0.618 &  3.083 &  0.946 & -0.474 & -0.259 & -0.284 &  0.031 &  0.155 & 16.7$^{\rm e}$ & T7.5 & Gl 570D \\
  4.261 &  0.582 &  3.093 &  1.061 & -0.433 & -0.152 & -0.296 &  0.034 &  0.152 & 17.0$^{\rm c}$ & T8 & 2MASS J04151954-0935066 \\
\hline
\end{tabular}
\begin{minipage}{165mm}
a Dahn et al. (2002)

b val Altena, Lee \& Hoffleit (1995)

c Vrba et al. (2004)

d Tinney et al. (2003)

e ESA (1997)
\end{minipage}
\end{scriptsize}
\end{table*}

\subsubsection{L and T dwarfs}

Colours of 30 L dwarfs and 22 T dwarfs have been computed from spectra
covering the wavelength range $i$ to $K$, collated from the following
references: Burgasser et al.  (2003), Chiu et al.  (2006),
Cruz et al.  (2003), Fan et al.  (2000a), Geballe et al.
(1996), Geballe et al.  (2001), Geballe et al (2002), Kirkpatrick,
Beichman \& Skrutskie (1997), Kirkpatrick et al.  (1999b), Kirkpatrick
et al.  (2000), Knapp et al.  (2004), Leggett et al. (2000; 2001;
2002a; 2002b), Liebert et al.  (2000), McLean et al.  (2003), Reid et
al.  (2001), Schultz et al.  (1998), Strauss et al.  (1999), Tinney et
al.  (1998), and Tsvetanov et al.  (2000).  In Table \ref{landt}, cols
$1-9$ provide colours over the range $i$ to $K$, col.  10 gives the
$K$-band absolute magnitude, M$_K$, with the source of the parallax
determination indicated, col. 11 provides the spectral classification
and col.  12 lists the object name.  As with the M dwarfs, in a few
cases the spectrum does not cover the entire $i$ band, and the $i-z$
entry is consequently blank.  The accuracy of the synthetic colours
that include the $i$, $Z$, $Y$ bands is subject to the same remarks
made above in relation to the M dwarfs.

The colours of very cool stars, of spectral type M, L, and T, across
the WFCAM bands, deserve comment.  In the $J-H$ {\it vs} $H-K$
two-colour diagram, Fig.  \ref{fig2colourstars}, moving down the brown
dwarf sequence, the early L stars firstly extend the O to M colour
sequence to redder colours.  Moving to cooler temperatures the colours
reverse and track back up the colour sequence, becoming successively
bluer, until the coolest T dwarfs are bluer than O stars (due to the
presence of strong absorption bands of water, methane and
pressure-induced H$_2$; see e.g.  Burgasser et al.  2002; Geballe et
al.  2002).  The consequence is that over a broad temperature range,
from mid L to mid T, the $J-H$ {\it vs} $H-K$ two--colour diagram is
ineffective for distinguishing brown dwarfs from hydrogen burning
stars.  A two-colour diagram involving $Z$, such as $Z-J$ {\it vs}
$J-H$ separates the two populations.  Nevertheless very deep
observations would be required in $Z$ to detect the coolest T dwarfs,
which have $Z-J\sim 4$.  The $Y$ band provides a compromise solution.
In the $Y-J$ {\it vs} $J-H$ diagram, plotted in the middle panels in
Fig.  \ref{fig2colourstars}, it is seen that brown dwarfs have
approximately constant $Y-J\sim 1.2$, redder than the M stars, while
the $J-H$ colour (or $J-K$) separates the L dwarfs from the T dwarfs.

\begin{table*}
\caption{Colours of Burrows et al. (2003) model T and Y cool dwarfs}
\label{burrows}
\centering
\begin{scriptsize}
\begin{tabular}{rrrrrrrrrrcccc}\\ \hline
\multicolumn{1}{c}{\iz} & \multicolumn{1}{c}{\Zz} & 
\multicolumn{1}{c}{\ZY} & \multicolumn{1}{c}{\YJ} & 
\multicolumn{1}{c}{\JH} & \multicolumn{1}{c}{\HK} & 
\multicolumn{1}{c}{\JJ} & \multicolumn{1}{c}{\HH} & 
\multicolumn{1}{c}{\KK} & \multicolumn{1}{c}{M$_K$} &
\multicolumn{1}{c}{mass} &\multicolumn{1}{c}{$\log t$} & 
\multicolumn{1}{c}{\tf} & \multicolumn{1}{c}{$\log g$} \\ 
 & & & & & & & & & & \multicolumn{1}{c}{$M_{\mathrm{Jup}}$} & \multicolumn{1}{c}{yr} &
\multicolumn{1}{c}{K} & \multicolumn{1}{c}{${\rm cm}\,{\rm s}^{-2}$} \\ 
\hline
4.215 & 2.247 & 5.829 & 2.464 & -0.729 & 0.782 & -0.277 & 0.070 & 0.100 & 23.3 & 1 & 8.0 & 290 & 3.23 \\
8.595 & 4.556 & 8.391 & 4.345 & 1.175 & 0.667 & -0.323 & 0.077 & 0.103 & 31.3 & 1 & 8.5 & 216 & 3.27 \\
11.662 & 6.062 & 9.928 & 5.510 & 2.278 & -3.478 & -0.377 & 0.082 & 0.055 & 43.5 & 1 & 9.0 & 159 & 3.32 \\
3.238 & 1.000 & 4.193 & 1.548 & -1.018 & 0.194 & -0.272 & 0.073 & 0.110 & 20.0 & 2 & 8.0 & 386 & 3.53 \\
4.162 & 2.130 & 5.626 & 2.349 & -0.529 & -0.117 & -0.293 & 0.074 & 0.105 & 24.4 & 2 & 8.5 & 283 & 3.57 \\
8.348 & 4.314 & 8.043 & 4.105 & 1.325 & -1.464 & -0.347 & 0.080 & 0.109 & 33.8 & 2 & 9.0 & 208 & 3.60 \\
12.773 & 5.796 & 9.630 & 6.817 & 3.962 & -2.575 & -0.345 & 0.072 & -0.121 & 45.8 & 2 & 9.5 & 149 & 3.63 \\
13.659 & 5.121 & 8.968 & 8.828 & 5.572 & -2.164 & -0.117 & -0.088 & -0.514 & 49.2 & 2 & 9.7 & 134 & 3.64 \\
4.782 & 0.612 & 3.530 & 1.264 & -0.696 & 0.129 & -0.289 & 0.074 & 0.133 & 16.9 & 5 & 8.0 & 588 & 3.92 \\
4.417 & 0.835 & 3.898 & 1.509 & -1.134 & -0.448 & -0.290 & 0.078 & 0.116 & 19.7 & 5 & 8.5 & 426 & 3.96 \\
3.338 & 1.510 & 4.767 & 1.879 & -0.414 & -1.031 & -0.302 & 0.077 & 0.112 & 23.7 & 5 & 9.0 & 312 & 3.99 \\
7.888 & 3.889 & 7.526 & 4.021 & 1.910 & -1.984 & -0.350 & 0.083 & 0.117 & 32.5 & 5 & 9.5 & 225 & 4.02 \\
10.624 & 5.149 & 8.955 & 5.397 & 3.329 & -1.306 & -0.357 & 0.094 & 0.112 & 36.7 & 5 & 9.7 & 197 & 4.03 \\
4.661 & 0.443 & 3.062 & 1.288 & -0.454 & 0.114 & -0.292 & 0.071 & 0.137 & 15.8 & 7 & 8.0 & 703 & 4.07 \\
4.715 & 0.769 & 3.778 & 1.487 & -0.833 & -0.261 & -0.291 & 0.080 & 0.125 & 18.2 & 7 & 8.5 & 507 & 4.12 \\
2.923 & 0.888 & 3.860 & 1.468 & -0.592 & -0.998 & -0.294 & 0.078 & 0.115 & 21.6 & 7 & 9.0 & 369 & 4.15 \\
4.418 & 2.109 & 5.423 & 2.318 & 0.326 & -2.213 & -0.327 & 0.078 & 0.116 & 27.3 & 7 & 9.5 & 267 & 4.18 \\
5.737 & 2.799 & 6.199 & 2.905 & 0.982 & -3.279 & -0.345 & 0.078 & 0.119 & 30.8 & 7 & 9.7 & 234 & 4.19 \\
4.190 & 0.301 & 2.457 & 1.409 & -0.147 & 0.092 & -0.288 & 0.074 & 0.132 & 14.6 & 10 & 8.0 & 859 & 4.24 \\
5.044 & 0.646 & 3.501 & 1.398 & -0.553 & -0.159 & -0.292 & 0.077 & 0.134 & 16.8 & 10 & 8.5 & 620 & 4.29 \\
4.491 & 0.787 & 3.733 & 1.638 & -0.892 & -0.877 & -0.298 & 0.081 & 0.120 & 19.7 & 10 & 9.0 & 447 & 4.32 \\
3.621 & 0.690 & 3.364 & 1.035 & -1.054 & -2.160 & -0.318 & 0.076 & 0.118 & 24.1 & 10 & 9.5 & 325 & 4.35 \\
3.848 & 1.730 & 4.914 & 2.042 & 0.297 & -2.128 & -0.326 & 0.078 & 0.120 & 26.1 & 10 & 9.7 & 284 & 4.36 \\
4.924 & 0.703 & 3.555 & 1.546 & -0.539 & -0.446 & -0.294 & 0.081 & 0.130 & 17.3 & 15 & 9.0 & 593 & 4.52 \\
4.117 & 0.680 & 3.448 & 1.582 & -0.854 & -1.502 & -0.307 & 0.081 & 0.122 & 21.0 & 15 & 9.5 & 414 & 4.56 \\
2.807 & 0.802 & 3.568 & 1.409 & -0.197 & -1.903 & -0.309 & 0.080 & 0.122 & 22.8 & 15 & 9.7 & 359 & 4.57 \\
5.172 & 0.640 & 3.379 & 1.475 & -0.390 & -0.367 & -0.293 & 0.076 & 0.132 & 16.3 & 20 & 9.0 & 686 & 4.66 \\
4.465 & 0.709 & 3.493 & 1.732 & -0.661 & -1.190 & -0.302 & 0.085 & 0.125 & 19.4 & 20 & 9.5 & 483 & 4.71 \\
4.095 & 0.642 & 3.323 & 1.584 & -0.752 & -1.671 & -0.310 & 0.083 & 0.124 & 21.0 & 20 & 9.7 & 421 & 4.72 \\
5.107 & 0.516 & 3.015 & 1.452 & -0.269 & -0.295 & -0.290 & 0.073 & 0.130 & 15.4 & 25 & 9.0 & 797 & 4.80 \\
4.762 & 0.733 & 3.517 & 1.747 & -0.517 & -0.915 & -0.298 & 0.085 & 0.126 & 18.2 & 25 & 9.5 & 555 & 4.83 \\
4.578 & 0.698 & 3.429 & 1.752 & -0.624 & -1.344 & -0.304 & 0.086 & 0.125 & 19.6 & 25 & 9.7 & 483 & 4.85 \\
\hline
\end{tabular}
\end{scriptsize}
\end{table*}

\begin{figure}
\begin{center}
\includegraphics[width=9cm]{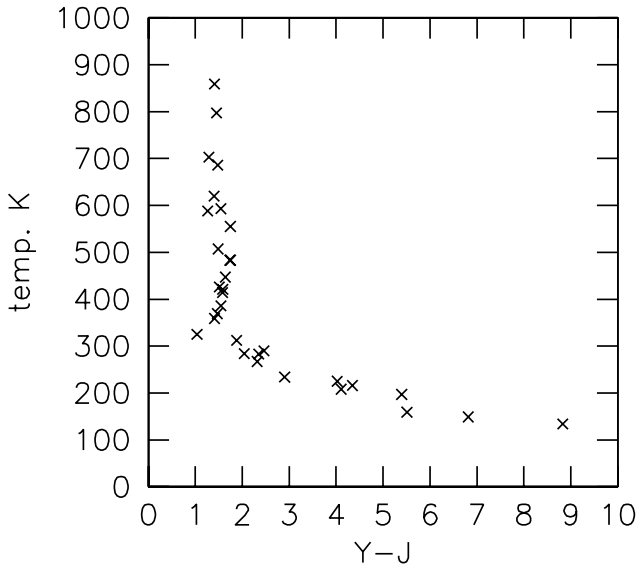}
\caption{\label{figtyj}Temperature--colour relation of
    model cool brown dwarfs}
\end{center}
\end{figure}

\subsubsection{Model Y dwarfs}
\label{sec:ydwarfs}

The current coolest known T dwarf is 2MASSJ0415-0935 (Burgasser et al.  2002).
The synthetic colours for this object are included in Table \ref{landt}.  The
star has \tf$\sim 700$K (Golimowski et al.  2004), and spectral type T9 (Knapp
et al, 2004) in the Geballe et al.  (2002) classification scheme, revised to T8
by Burgasser, Burrows \& Kirkpatrick (2006).  Cooler brown dwarfs no doubt
exist, but will be difficult to find, because of their very low luminosities.
One of the main goals of UKIDSS is to discover such stars.  At some temperature
a new spectral sequence is expected to appear, possibly associated with the
emergence of NH$_3$ in the spectrum, and the nomenclature Y dwarf has been
suggested (Kirkpatrick et al.  1999a, Kirkpatrick 2000).  In order to develop a
strategy for finding brown dwarfs cooler than \tf$\sim 700$K we have computed
synthetic colours from the model spectra of Burrows, Sudarsky \& Lunine (2003),
which cover masses in the range $1-10M_{\mathrm{Jup}}$, and ages
$10^{8.0-9.7}$yr.  In Table \ref{burrows}, cols $1-9$ provide predicted colours
over the range $i$ to $K$, col.  10 the $K$-band absolute magnitude, M$_K$, and
cols $11-14$ list, respectively, the model mass, age, \tf, and surface gravity.

\begin{table*}
\caption{Colours of Marley et al. model cool dwarfs}
\label{marley}
\centering
\begin{scriptsize}
\begin{tabular}{rrrrrrrrrrcc}\\ \hline
\multicolumn{1}{c}{\iz} & \multicolumn{1}{c}{\Zz} & 
\multicolumn{1}{c}{\ZY} & \multicolumn{1}{c}{\YJ} & 
\multicolumn{1}{c}{\JH} & \multicolumn{1}{c}{\HK} & 
\multicolumn{1}{c}{\JJ} & \multicolumn{1}{c}{\HH} & 
\multicolumn{1}{c}{\KK} & \multicolumn{1}{c}{M$_K$} &
\multicolumn{1}{c}{\tf} & \multicolumn{1}{c}{$\log g$} \\ 
 & & & & & & & & & &
\multicolumn{1}{c}{K} & \multicolumn{1}{c}{${\rm cm}\,{\rm s}^{-2}$} \\ \hline
   6.027 &  0.651 &  3.467 &  0.849 & -1.093 & -1.034 & -0.294 &  0.072 &  0.135 & 18.2 & 600 & 4.48 \\
   5.848 &  0.694 &  3.359 &  0.937 & -0.919 & -1.779 & -0.296 &  0.074 &  0.131 & 18.7 & 600 & 5.00 \\
   5.806 &  0.730 &  3.269 &  0.970 & -0.680 & -1.762 & -0.294 &  0.073 &  0.125 & 18.7 & 600 & 5.48 \\
   5.908 &  0.560 &  3.237 &  0.942 & -0.854 & -0.749 & -0.289 &  0.067 &  0.136 & 16.9 & 700 & 4.48 \\
   5.942 &  0.666 &  3.287 &  0.985 & -0.786 & -1.279 & -0.291 &  0.070 &  0.129 & 17.3 & 700 & 5.00 \\
   5.976 &  0.779 &  3.341 &  1.031 & -0.673 & -1.402 & -0.288 &  0.068 &  0.121 & 17.5 & 700 & 5.48 \\
\hline
\end{tabular}
\begin{minipage}{165mm}
Note: the model spectra extend only to $2.4\,\mu$m and the $H-K$ and $K-K2$
colours may be in error by up to $0.02\,$mag.
\end{minipage}
\end{scriptsize}
\end{table*}

Table \ref{marley} lists colours computed from similar models by
Marley et al.  (2002; 2006, in preparation).  Cols 1-10 in Table 12
contain the identical information to that for the Burrows et al.
models in Table 11, while cols 11 and 12 specify the temperature and
surface gravity for the Marley et al.  models.  These models are for
T$_{\rm eff}$ of 700 and 600\,K, and surface gravities, ${\rm log}\,g$ of
4.48, 5.00 and 5.48$\,{\rm cm}\,{\rm s}^{-2}$, and so sample masses
around 15, 30 and 60 $M_{\mathrm{Jup}}$ aged $0.4-0.6$, $3.0-4.6$ and 
$\geq$
13 Gyr, respectively.  Although the $Y-J$ colours of the Marley models
are bluer than the Burrows models, the general trends are similar ---
very late T and Y dwarfs will have $Y-J$ redder than M dwarfs ($Y-J >
0.8)$ while $J-H$ will be extremely blue ($J-H < -0.2$).

The synthetic colours for the models of warmer temperature, which
overlap in temperature with the coolest known T dwarfs, are in
reasonable agreement with the measured colours of the T dwarfs (Fig.
\ref{fig2colourstars}), giving some confidence in the synthetic
colours of the cooler objects.  The range of masses and ages
considered give rise to a wide range of predicted colours.  The models
however all suggest that the coolest brown dwarfs continue getting
bluer in the near-infrared with decreasing T$_{\rm eff}$ for T$_{\rm
eff} > 400\,$K.  Burrows et al. (2003) describe the
reddening at cooler temperatures to be due to the appearance of water
clouds and, more importantly, the collapse of flux on the Wien tail.

Spectral changes occur at temperatures hotter than 400\,K: the alkali
lines are expected to disappear below 500\,K, and at 600\,K NH$_3$ is
expected to be detectable at the blue edge of the $H$ and $K$ band
peaks (Burrows et al.  2003).  The appearance of NH$_3$ in the
near-infrared may signal the next spectral type after T, i.e. Y, as
discussed above.  WFCAM will allow candidate very late T and Y dwarfs
to be identified from their blue $J-H$ colours; these objects will be
followed up spectroscopically.  The LAS should find several brown
dwarfs later than T8 in the first two years of the survey.  Adopting a
detection limit of $K=18.4$, the LAS should detect dwarfs as cool as
450\,K at 10\,pc in all $YJH$ bands (col.  10 in Tables 11 and 12).  This
temperature limit translates to a lower mass limit of 10\,$M_{\mathrm {Jup}}$
for an assumed age of 1-5\,Gyr.

Should young, nearby, even cooler objects be detected in the $Y$ and
$J$ bands, Fig.  \ref{figtyj} shows that the model $Y-J$ colours are
approximately constant with temperature down to 400K, then move
rapidly redder.  Stellar objects redder than $Y-J=2$ would be
extremely cool, and very interesting objects indeed.  Possible
contaminants include carbon stars, distinguishable using other
colours, and quasars of very--high redshift, $z>7.8$, which are
expected to be extremely rare.  Although model colours of elliptical
galaxies become very red in $Y-J$ at high redshift $z>2.8$ (see
below), this is only true of an unevolving spectrum.  For realistic
formation redshifts, such extreme colours would not be seen.  For the
model cool brown dwarfs, a similar trend with temperature is seen in
the $Z-J$ colour, but a selection limit $Z-J>5$ would be required,
which would involve unreasonably long integrations in the $Z$ band.
Caution is required however as the model colours are subject to
uncertainty, and other authors (Baraffe et al. 2003; Marley, private
communication) find somewhat different trends.

\begin{table*}
\caption{Colours of pure H atmosphere model white dwarf spectra of
  different effective temperature}
\label{bergeronh}
\centering
\begin{scriptsize}
\begin{tabular}{rrrrrrrrrrrrrr}\\ \hline
\multicolumn{1}{c}{\ug} & \multicolumn{1}{c}{\gr} &
\multicolumn{1}{c}{\ri} & \multicolumn{1}{c}{\iz} & 
\multicolumn{1}{c}{\Zz} & \multicolumn{1}{c}{\ZY} & 
\multicolumn{1}{c}{\YJ} & \multicolumn{1}{c}{\JH} & 
\multicolumn{1}{c}{\HK} & \multicolumn{1}{c}{\JJ} & 
\multicolumn{1}{c}{\HH} & \multicolumn{1}{c}{\KK} & 
\multicolumn{1}{c}{$M_K$} & \multicolumn{1}{c}{\tf} \\
 & & & & & & & & & & & & & \multicolumn{1}{c}{K} \\
\hline
  1.632 &  0.932 & -2.161 & -0.054 & -0.147 & -1.709 &  0.832 &  0.467 & -2.569 & -0.045 & -0.139 & -0.036 & 23.937 &  1500  \\
  1.410 &  1.204 & -1.555 & -0.076 & -0.124 & -1.496 &  0.747 &  0.243 & -1.871 & -0.057 & -0.107 & -0.032 & 22.053 &  1750  \\
  1.238 &  1.334 & -1.008 & -0.106 & -0.105 & -1.313 &  0.668 &  0.072 & -1.341 & -0.058 & -0.083 & -0.033 & 20.559 &  2000  \\
  1.091 &  1.391 & -0.517 & -0.128 & -0.091 & -1.152 &  0.587 & -0.061 & -0.963 & -0.054 & -0.066 & -0.034 & 19.380 &  2250  \\
  0.962 &  1.401 & -0.092 & -0.113 & -0.078 & -0.998 &  0.507 & -0.154 & -0.681 & -0.046 & -0.051 & -0.031 & 18.390 &  2500  \\
\hline
\end{tabular}
\begin{minipage}{165mm}
Note: The full table is published in the electronic version of the
paper. A portion is shown here for guidance regarding its form and content.
\end{minipage}
\end{scriptsize}
\end{table*}

\begin{table*}
\caption{Colours of pure He atmosphere model white dwarf spectra of
  different effective temperature}
\label{bergeronhe}
\centering
\begin{scriptsize}
\begin{tabular}{rrrrrrrrrrrrrr}\\ \hline
\multicolumn{1}{c}{\ug} & \multicolumn{1}{c}{\gr} &
\multicolumn{1}{c}{\ri} & \multicolumn{1}{c}{\iz} & 
\multicolumn{1}{c}{\Zz} & \multicolumn{1}{c}{\ZY} & 
\multicolumn{1}{c}{\YJ} & \multicolumn{1}{c}{\JH} & 
\multicolumn{1}{c}{\HK} & \multicolumn{1}{c}{\JJ} & 
\multicolumn{1}{c}{\HH} & \multicolumn{1}{c}{\KK} & 
\multicolumn{1}{c}{$M_K$} & \multicolumn{1}{c}{\tf} \\
 & & & & & & & & & & & & & \multicolumn{1}{c}{K} \\
\hline 
  1.994 &  2.123 &  1.111 &  0.691 &  0.028 &  0.394 &  0.476 &  0.444 &  0.316 & -0.029 &  0.013 & -0.016 & 13.858 &  3500  \\
  1.723 &  1.915 &  0.975 &  0.591 &  0.022 &  0.326 &  0.415 &  0.388 &  0.276 & -0.026 &  0.011 & -0.014 & 13.710 &  3750  \\
  1.464 &  1.715 &  0.851 &  0.504 &  0.017 &  0.268 &  0.363 &  0.341 &  0.243 & -0.024 &  0.010 & -0.013 & 13.583 &  4000  \\
  1.211 &  1.526 &  0.741 &  0.430 &  0.013 &  0.219 &  0.320 &  0.301 &  0.216 & -0.021 &  0.009 & -0.011 & 13.471 &  4250  \\
  0.960 &  1.346 &  0.645 &  0.368 &  0.009 &  0.178 &  0.284 &  0.269 &  0.192 & -0.020 &  0.008 & -0.010 & 13.373 &  4500  \\
\hline
\end{tabular}
\begin{minipage}{165mm}
Note: The full table is published in the electronic version of the
paper. A portion is shown here for guidance regarding its form and content.
\end{minipage}
\end{scriptsize}
\end{table*}

\subsubsection{Cool white dwarfs}

Model spectra of white dwarfs have been provided by P.  Bergeron (see
e.g.  Bergeron et al.  2005, references therein, and the web page {\tt
http://www.astro.umontreal.ca/$\sim$bergeron/CoolingModels/})
and used to calculate colours over the range $u$ to $K$.  The models
are for pure H and pure He atmospheres, with surface gravity log\,$g=$8.
Tables \ref{bergeronh} and \ref{bergeronhe} provide the colours in
columns $1-12$, column 13 provides the $K-$band absolute magnitude,
and column 14 the effective temperature.

Fig.  \ref{fig2colourstars} shows that while He--atmosphere white dwarfs are
difficult to distinguish from main sequence stars, hydrogen--rich white dwarfs
become blue in $J-H$ at T$_{\rm eff}<$5000\,K.  This is due to the onset of
pressure--induced molecular hydrogen absorption in these high--pressure
atmospheres.  White dwarfs can be used as chronometers, as their cooling after
the planetary nebula stage is reasonably well understood.  The age of the
Galactic disk has been constrained by the coolest disk white dwarfs (Liebert,
Dahn \& Monet 1988; Oswalt et al.  1996), and if halo white dwarfs can be
identified, the age of the halo could also be constrained.  This is especially
true of the H--rich white dwarfs, as He atmospheres are less opaque and allow
the white dwarf to cool rapidly \---\ a He--rich dwarf will be a Gyr or more
younger than a H--rich dwarf with the same temperature (e.g.  Bergeron, Leggett
\& Ruiz 2001).  Fig.  9 of Bergeron et al.  (2005) shows that most white dwarfs
known have T$_{\rm eff}>$4000\,K and, based on tangential velocity and
temperature, are most likely to be (thick) disk dwarfs younger than 10$^{10}$
years.  The LAS will be able to detect 4000\,K H--rich white dwarfs out to 
around 60\,pc, and 3000\,K H--rich white dwarfs to $20\,$pc.  While these 
objects are very
rare, they occupy a unique region of the $J-H$ {\it vs} $Y-J$ diagram and will
be easy to identify and important to follow-up.

\subsection{Galaxies: {\em hyperz} templates, and the Kinney--Mannucci
    spectra}

We have computed synthetic colours of galaxies over the redshift range
$0<z<3.6$, with a step size $\Delta z=0.1$, by redshifting unevolving
locally-measured template spectra of different galaxy types.
Predictions for the $u$ and $g$ colours at high redshift require the
adoption of a model for the effects of intervening absorption.  While
such a model is employed to make predictions for the quasar SEDs
(Section 5.4), galaxy colours at high redshift based on unevolved
ultraviolet galaxy SEDs are of limited utility and the $u$- and
$g$-band simulations are confined to redshifts $z \le 1.7$ and $\le
2.5$ respectively.  We have used two published sets of templates, as
detailed below.

\begin{table}
\caption{List of Tables containing predicted galaxy and quasar colours}
\label{tab_desc}
\centering
\begin{scriptsize}
\begin{tabular}{ll}\\ \hline
\multicolumn{1}{c}{Table No.} & Description \\ \hline 
Table 16 & Colours of redshifted {\em hyperz} E galaxy\\
Table 17 & Colours of redshifted {\em hyperz} Sbc galaxy\\
Table 18 & Colours of redshifted {\em hyperz} Scd galaxy\\
Table 19 & Colours of redshifted {\em hyperz} Im galaxy\\
Table 20 & Colours of redshifted Kinney--Mannucci E galaxy\\
Table 21 & Colours of redshifted Kinney--Mannucci S0 galaxy\\
Table 22 & Colours of redshifted Kinney--Mannucci Sa galaxy\\
Table 23 & Colours of redshifted Kinney--Mannucci Sb galaxy\\
Table 24 & Colours of redshifted Kinney--Mannucci Sc galaxy\\
Table 25 & Colours of redshifted blue model quasar\\
Table 26 & Colours of redshifted average model quasar\\
Table 27 & Colours of redshifted red model quasar\\ 
\hline
\end{tabular}
\end{scriptsize}
\end{table}

\subsubsection{hyperz templates}

The first set of templates are the extended Coleman, Wu \& Weedman
(1980) spectra that are supplied with and used by the {\em hyperz}
photometric--redshift code (Bolzonella, Miralles \& Pell{\' o}, 2000).
The original Coleman et al.  spectra cover the wavelength range $1400
< \lambda < 10\,000$ \AA, and were extended by Bolzonella et
al.  to both shorter and longer wavelengths using synthetic spectra
created with the GISSEL98 code (Bruzual \& Charlot, 1993).  The
synthetic colours for, respectively, the template E, Sbc, Scd and Im
spectra are provided in Tables 16-19.  In
each table col.  1 lists the redshift, and cols $2-12$ list colours in
the $u$ to $K$ bands.  Blank entries correspond to redshift ranges
where colours would be affected by intergalactic absorption. Table
\ref{tab_desc} summarises the contents of the tables containing the
predicted colours for the various galaxy and quasar template spectra.

The redshift tracks of the $Z$$Y$$J$$H$$K$ colours of the four spectra
are plotted in Fig.  \ref{fig3colourgalaxies}.  The effect of redshift
is, as a rule, to shift the spectra redward of the locus of stars,
i.e.  towards the bottom right in each plot.  We chose to redshift
unevolving templates as this gives an idea of the envelope within
which galaxy colours lie at any redshift.  But the use of an old
stellar population (the E spectrum) to define the red boundary of the
colour distribution, becomes unrealistic by redshift $z=2$.  Therefore
objects discovered in the region of colour space traced by the E
spectrum at high redshift, are more likely to be dusty galaxies, not
represented by the templates used here.

\subsubsection{Kinney--Mannucci spectra}

The second set of templates are the spectra created by Mannucci et al.
(2001), by combining their near-infrared averaged spectra with the
UV--optical template spectra of Kinney et al.  (1996).  The synthetic
colours for, respectively, the template E, S0, Sa, Sb, and Sc spectra
are provided in Tables 20-24.  In each table col.  1
lists the redshift, and cols $2-12$ list colours in the $u$ to $K$
bands.  Blank entries correspond to redshift ranges where colours
would be affected by intergalactic absorption or where the template
SEDs provide incomplete coverage.  The latter restriction is confined
to colours involving $K$ at $z=0$ as the spectra of Mannucci et al.
(2001) reach only $2.4\,\mu$m, where the WFCAM $K$ filter still has
significant transmission.

The redshift tracks of the Kinney--Mannucci spectra mostly follow
closely the {\em hyperz} tracks for similar spectral classes.

\begin{figure}
\begin{center}
\includegraphics[width=9cm]{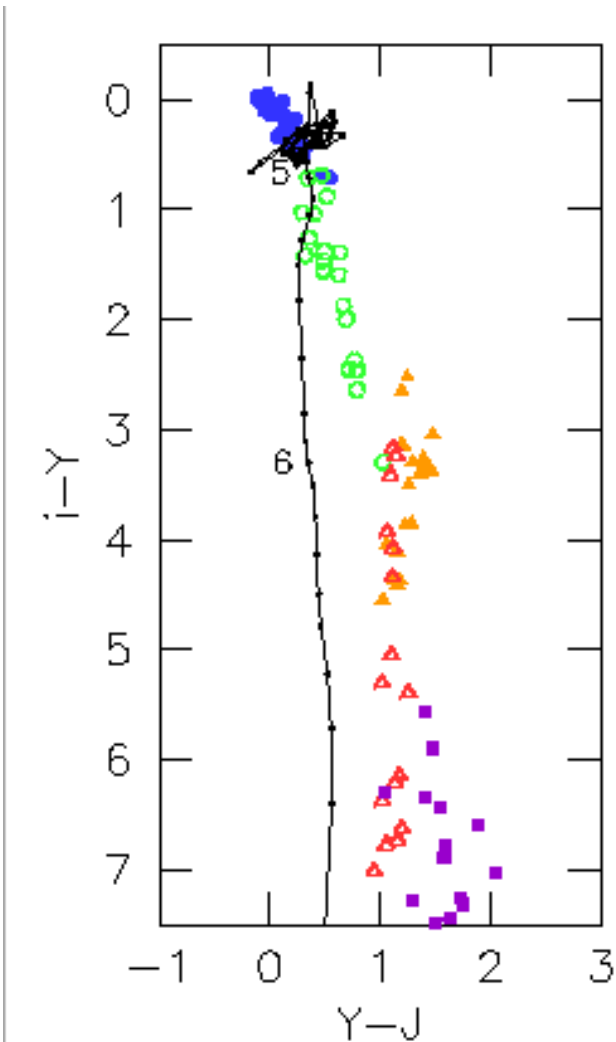}
\caption{\label{fig5colourquasars}2--colour diagram illustrating
     colours of stars, brown dwarfs and quasars.  Key: BPGS O--K dwarfs
     blue {\Large \textbullet}; M dwarfs green $ \bigcirc$; L dwarfs
     orange $\blacktriangle$; T dwarfs red $\triangle$; Burrows model
     cool brown dwarfs purple $\blacksquare$; quasars $0<z<8.5$, $\Delta
     z=0.1$, solid black line, with redshifts 5 and 6 marked. In this 
plot
     quasars $6<z<7.2$ are identifiable as bluer than L and T dwarfs in
     the $Y-J$ colour. Note that the Marley model cool brown dwarfs all
     have $i-Y>7.5$ and so lie off the plot.}
\end{center}
\end{figure}

\subsection{Quasars in the redshift range $0<z<8.5$}
\label{sec:quasars}

Given the extended wavelength range covered by the passbands ($u$
through $K$) and the large redshift range of interest for the quasars,
$0 < z < 8.5$, a quasar SED covering the rest-frame wavelength range
$500 < \lambda < 25\,000$ \AA \ is required.  High
signal-to-noise ratio composite spectra from large quasar surveys
(Francis et al.  1991; Brotherton et al.  2001; Vanden Berk et al.
2001) provide information for the wavelength range $1000 \la \lambda
\la 7000$ \AA \ but additional constraints are needed to model
the full wavelength range.  The relatively simple model SED employed
here has been constructed using the emission line properties of the
Large Bright Quasar Survey (LBQS) composite spectrum (Francis et al.
1991) combined with a parametric continuum model.  A more complete
description of the quasar model SED and the data used to determine the
parameters is given in Maddox \& Hewett (2006) but an
outline is given here.

The quasar SED model parameters were defined by requiring the model to
reproduce the median colours in $ugrizJ2H2K2$ of a sample of $2708$
quasars, magnitudes $i < 17.4$ and redshifts $0.1 \le z \le
3.6$, with unresolved morphologies, from the SDSS DR3 quasar catalogue
(Schneider et al.  2005), 90\% of which possess 2MASS $JHK$
detections.  The ``continuum'' is represented by a power-law in
frequency, $F(\nu) \propto \nu^\alpha$ with $\alpha = -0.3$ for
$\lambda < 12\,000\,$\AA \ and $\alpha = -2.4$ for $\lambda \ge
12\,000\,$\AA.  A modest contribution from Balmer continuum radiation
is included to model the excess flux needed at $\sim 3000\,$\AA.
Emission line contributions from the LBQS composite spectrum (Francis
et al.  1991) over the wavelength range $1000-6000\,$\AA \ are
included, along with H$\alpha$ from an extended version of the LBQS
composite, and the addition of a Paschen-$\alpha$ line.  The H$\alpha$
strength in the composite is derived from low-luminosity objects and
it is necessary to incorporate a reduction in the equivalent width of
H$\alpha$ as a function of luminosity, i.e.  a Baldwin Effect (Baldwin
1977), with a dependence $EW({\rm H}\alpha) \propto L_{qso}^{-0.2}$,
to reproduce the behaviour of the median $JHK$ colours as H$\alpha$
passes through the near-infrared passbands over the redshift range $1
\la z \la 3$.

Given the primary redshift range of interest for the UKIDSS survey is at $z>6$,
the effects of intergalactic absorption are incorporated by including the
absorption from an infinite optical-depth Lyman-limit system at $912\,$\AA \ in
the quasar restframe, i.e.  there is no flux below $912\,$\AA \ in the quasar
restframe.  Absorption due to Lyman-$\alpha$, $\beta$ and $\gamma$ transitions
in the Lyman-$\alpha$ forest are included using data from Songaila (2004), which
extends to a redshift of $z=6.5$.  At redshift $z=6.5$ there is almost no flux
shortward of $1216\,$\AA \ and rather than attempting an unconstrained
extrapolation of the absorption behaviour to higher redshifts the absorption
applicable at $z=6.5$ has been applied at all redshifts $z > 6.5$.  In practice
the adoption of the $z=6.5$ absorption for higher redshifts means that colours
involving $Z$ and $Y$ (e.g.  $Y-J$) will represent blue limits.  However, the
amplitude of the discontinuity at $1216\,$\AA \ is such that such colours are
already extremely red (reaching 8 magnitudes) and additional absorption will
simply increase the amplitude of the continuum breaks relevant to the detection
of quasars at $z > 6.5$. Those interested in the $u$ band colours of quasars 
at redshifts $1.8 < z < 3.5$ will likely wish to apply a more sophisticated
Monte-Carlo approach to simulating the effects of different sight-lines through
the intergalactic medium. A somewhat steeper power-law index below the 
Lyman-$\alpha$ line would also provide an improved fit to the median $u$ band
colours. 

The simple model quasar SED provides an exceptionally good fit to the median
colours, $grizJ2H2K2$, over the full redshift range $0.1 \le z \le 3.6$ and
colours involving the $u$ band to $z \sim 2$.  To provide an indication of the
effects of changing the overall shape of the quasar SED on the UKIDSS colours,
``blue'' and ``red'' quasar SEDs have been generated by modifying the power-law
slope at $\lambda < 12000\,$\AA \ to $\alpha = 0.0$ and $\alpha = -0.6$
respectively.

The model colours of the blue, average, and red model quasar spectra,
respectively, are provided in Tables 25, 26 and 27,
and cover the redshift range
$0<z<8.5$, with redshift interval $\Delta z=0.1$.  In each table col.
1 lists the redshift, and cols $2-13$ provide colours over the range
$u$ to $K$.

The discovery of quasars of very--high redshift $z>6$ is another of
the key goals of UKIDSS (Warren \& Hewett, 2002).  The general
principle used to find high--redshift quasars $3<z<6.4$ with broadband
photometry (e.g.  Warren, Hewett \& Osmer, 1991; Fan et al.  2000b) is
to image in three bands, one ($a$) blueward of redshifted Ly$\alpha$,
a second ($b$) containing or just redward of Ly$\alpha$, and a third
($c$) further to the red.  The quasars are then red in $a-b$, and
bluer in $b-c$ than any stars, or brown dwarfs, that are similarly red
in $a-b$.  The most distant quasar found in the SDSS has $z=6.40$ (Fan
et al.  2003; Iwamuro et al.  2004).  At higher redshifts Lyman-$\alpha$
moves out of the $z$ band, and a longer-wavelength middle filter is
needed.

Unfortunately in the standard near-infrared bands $J$$H$$K$ quasars
are redder than most stars (Warren, Hewett \& Foltz, 2000), and have
similar colours to M stars, late L dwarfs, and early T dwarfs.  This
means that in a search for quasars of redshift $z>6.4$ a colour from
the $J$$H$$K$ bands used as the red colour $b-c$ is not effective.  As
shown in Fig.  4, the $Y$ band appears to offer a solution.  The
colour of the average quasar spectrum lies at least 0.5mag.  bluer
than L and T dwarfs in $Y-J$ until $z\sim7.2$, so that combining
UKIDSS data with sufficiently deep $i-$band data it may be possible to
discover quasars at redshifts $z>6.4$.

\section{Colour equations between WFCAM, SDSS, and 2MASS}
\label{sec:colourterms}

We have used synthetic colours to compute colour equations relating photometry
in the WFCAM bands and the native SDSS 2.5m telescope $z$ band and 2MASS $J2$,
$H2$, and $K2$ bands i.e.  those bands with similar effective wavelengths to
WFCAM bands.  We used the BPGS atlas, and the additional M stars (Section
\ref{sec:mstars}), but excluded L and T dwarfs which follow different, more
complicated, relations.  We computed fits for luminosity classes III and V
separately.  The two fits are mostly similar for any band, but noticeably
different in the $K$ band.  The following relations allow photometric
transformations between the different systems.  The root mean square scatter in
these relations is $0.01\,$mag.  or less.  We remind readers that all magnitudes
here are Vega based, and that Table \ref{aboffsets} provides offsets to the 
AB system.

\( \begin{array}{rcllr}
     &   &    &  \\
Z & = & z-0.01+0.051(i-z) & \mathrm{[III]} & (2a) \\
Z & = & z-0.01+0.068(i-z) & \mathrm{[V]} & (2b) \\
z & = & Z+0.02-0.103(Z-Y) & \mathrm{[III]} & (2c) \\
z & = & Z+0.01-0.103(Z-Y) & \mathrm{[V]} & (2d) \\
     &   &    &  \\
J & = & J2-0.01-0.003(J2-H2) & \mathrm{[III]} & (3a) \\
J & = & J2+0.01-0.067(J2-H2) & \mathrm{[V]} & (3b) \\
J2 & = & J+0.02 & \mathrm{[III]} & (3c) \\
J2 & = & J-0.01+0.073(J-H) & \mathrm{[V]} & (3d) \\
     &   &    &  \\
H & = & H2+0.01+0.065(H2-K2) & \mathrm{[III]} & (4a) \\
H & = & H2+0.080(H2-K2) & \mathrm{[V]} & (4b) \\
H2 & = & H-0.01-0.063(H-K) & \mathrm{[III]} & (4c) \\
H2 & = & H-0.069(H-K) & \mathrm{[V]} & (4d) \\
     &   &    &  \\
K & = & K2+0.075(H2-K2) & \mathrm{[III]} & (5a) \\
K & = & K2-0.081(H2-K2) & \mathrm{[V]} & (5b) \\
K2 & = & K-0.072(H-K) & \mathrm{[III]} & (5c) \\
K2 & = & K+0.073(H-K) & \mathrm{[V]} & (5d)
\end{array} \)

\section*{Acknowledgments} 
The referee, Dr Allyn Smith, provided a careful and detailed reading of the
paper for which we are indebted.  We are grateful to Mark Casali for providing
the WFCAM filter transmission curves, to Xiaohui Fan for forwarding the SDSS
response functions, to Tom Geballe for providing the ATRAN results, to Pierre
Bergeron for the model cool white dwarf spectra and to Martin Cohen
for correspondence on photometric zero points for Vega.

\end{document}